# Multiplicity Distributions of Shower Particles and Target Fragments in $^{84}Kr_{36}$ - Emulsion Interactions at 1 GeV per nucleon

M K Singh[1, 2], A K Soma[1], Ramji Pathak[2], V Singh[*1]
1. Physics Department, Banaras Hindu University, Varanasi – 221005, India
2. Physics Department, Tilak Dhari Postgraduate College, Jaunpur – 222002, India
*Email: venkaz@yahoo.com

**Abstract**

The article focuses on the multiplicity distributions of shower particles and target fragments for the interaction of $^{84}Kr_{36}$ possessing kinetic energy of ~1 GeV per nucleon with NIKFI BR-2 Nuclear Emulsion target. The experimental multiplicity distributions of shower particles, grey fragments, black fragments, and heavily ionization fragments are well described by the multi-component Erlang distribution of the multi-source thermal model. We observed a linear correlation in multiplicities for the above mentioned particles or fragments. The further experimental studies showed that with the increase of target fragment multiplicity, a saturation phenomenon was observed in shower particle multiplicity.

## 1. Introduction

The investigation of final state particles produced in nucleon-nucleus or nucleus-nucleus interactions at high energy is an active research area with vast discovery potential [1-4]. Nuclear Emulsion Detector (NED) is one of the oldest detector technologies and has been in use from the birth of experimental nuclear and astroparticle physics. The high angular resolution of NED's allows easy detection of short lived particles like tau lepton, charmed meson etc.,. The size and range of ionization sensitivity of NED depends upon nature and need of the experiments. The interaction of nucleons (such as $^{84}Kr_{36}$) with NED's target at relativistic energy reveals the picture of the hadrons – nucleus collisions [3-8]. The recoil target nucleons are emitted shortly after the passage of leading hadrons. Therefore, it is worthy to study the multiplicity of recoil target fragments such as black fragments (b), grey fragments (g) and shower (s) particles produced during the interactions.

## 2. Experimental details

A stack of high sensitive NIKFI BR-2 nuclear emulsion pellicles of dimensions 9.8×9.8×0.06 cm$^3$ were exposed horizontally by $^{84}Kr_{36}$ ion of kinetic energy ~1 GeV per nucleon [3-6]. The exposure was performed at Gesellschaft fur Schwerionenforschung (GSI) Darmstadt, Germany. Interactions were found by along the track scanning technique using an oil immersion objective of 100X magnification. The beam tracks were picked up at a distance of 5 mm from the edge of plate and carefully followed until they either interacted with nuclear emulsion detector nuclei or escaped from any surface of the emulsion [4-6].

All charge secondary emitted or produced in an interaction are classified in accordance with their ionization or normalized grain density (g*), range (L) and velocity (β) into the following categories [5-8]:

*2.1 Shower tracks ($N_s$):* These are freshly created charged particles with g* less than 1.4. These particles have β > 0.7. They are mostly fast pions with a small mixture of Kaons and released protons from the projectile which have undergone an interaction. For the case of proton, kinetic energy ($E_p$) should be less than 400 MeV.

*2.2 Grey tracks ($N_g$):* Particles with range L > 3 mm and 1.4 < g* < 6.0 are defined as greys. They have β in the range of 0.3 < β < 0.7. These are generally knocked out protons (NED's can not detect neutral particle) of targets with kinetic energy in between 30 - 400 MeV, and traces of deuterons, tritons and slow mesons.

*2.3 Black tracks ($N_b$):* Particles having L < 3 mm from interaction vertex from and g* > 6.0. This corresponds to β < 0.3 and protons of kinetic energy less than 30 MeV. Most of these are produced due to evaporation of residual target nucleus.

The number of heavily (h) ionizing charged particles ($N_h$) are part of the target nucleus is equal to the sum of black and gray fragments ($N_h = N_b + N_g$).

## 3. The model

In this study, we used multi-source thermal model [9-13] to describe the characteristics of multiplicity distributions. In this model, many emission sources are assumed to form intermediate as well as high energy nucleus-nucleus collisions. In multiplicity distribution, the multi-source thermal model results a multi-component Erlang distribution which describes different kinds of particles as well as fragments [11-13]. On the basis of this model, we divided the experimental event sample into groups based on impact parameter and reaction mechanisms such as evaporation, absorption, spallation, multifragmentations etc.,.

Let us consider there are $\beta_j$ sources in the $j^{th}$ group. Each source is assumed to contribute an exponential form to the multiplicity distribution. Thus, we have the multiplicity ($\alpha_{ij}$) distribution contributed by the $i^{th}$ source in the $j^{th}$ group to be,

$$P_{ij}(\alpha_{ij}) = \frac{1}{<\alpha_{ij}>} \exp(\frac{\alpha_{ij}}{<\alpha_{ij}>}) . \qquad (1)$$

Where, ($\alpha_{ij}$) the mean multiplicity is contributed by the $i^{th}$ source in the $j^{th}$ group and is not related to i. The multiplicity distribution contributed by the $j^{th}$ group is an Enlang distribution,

$$P_j(\alpha_x) = \frac{\alpha_x^{\beta_j - 1}}{(\beta_j - 1)! <\alpha_{ij}>} \exp(\frac{\alpha_x}{<\alpha_{ij}>}) . \qquad (2)$$

It is the folding result of $\beta_j$ exponential functions and $\alpha_x = \sum_{i=1}^{\beta_j} \alpha_{ij}$, and x denotes s, g, b, h respectively. The multiplicity distribution obtained in final state is the weighed sum of l group contributions. We then have,

$$P(\alpha_x) = \frac{1}{N}\frac{dN}{d\alpha_x} = \sum_{j=1}^{l} k_j P_j(\alpha_x). \qquad (3)$$

Where, N and $k_j$ denote the particle fragment number and weight factor, respectively [11, 12].

In the Monte Carlo method, let $R_{ij}$ denote random numbers in (0, 1). Thus, equations (1) and (2) lead to, (4)

$$\alpha_{ij} = -<\alpha_{ij}\ln R_{ij},$$

and

$$\alpha_x = -\sum_{i=1}^{\beta_j} <\alpha_{ij}> \ln R_{ij}. \qquad (5)$$

The multiplicity distribution is then finally obtained by statistical method in accordance with different $k_j$ [9-12].

## 4. Results and Discussion

Figure 1, shows multiplicity distributions of Ns, $N_g$, $N_b$ and $N_h$ particles fragments in $^{84}Kr_{36}$ Emulsion collisions at ~1 GeV per nucleon. The histograms are experimental data while curves are modeling results. The mean multiplicities of $N_s$, $N_g$, $N_b$ and $N_h$ are 3.65 ± 0.07, 3.39 ± 0.09, 4.39 ± 0.24 and 5.74 ± 0.57, respectively. We used a two-component Erlang distribution for the calculation. The fitting parameter values are summarized n Table 1. We found that there are two different group, of events in the experimental data and the contributions from each groups is almost same ($k_1 = k_2 = 0.50 ± 0.10$). The first group corresponds to the light target nuclei (H / C / N / O) and peripheral collisions of projectile with heavy target nuclei (Ag / Br). The second group corresponds to Ag / Br. The mean multiplicities obtained from model as per the fitting details in Table 1 are 3.25, 3.47, 4.87, and 5.47 for Ns, $N_g$, $N_b$ and $N_h$, respectively. Thus, the experimental multiplicity values are well described by the model.

The model does not provide any information on correlations between Ns, $N_g$, $N_b$ and $N_h$ but we have observed a linear behavior in various correlations. Figures 2 demonstrates the correlations between $<N_s>$ - $N_g$, $<N_s>$ - $N_b$, $<N_s>$ - $N_h$, $<N_g>$ - $N_b$, $<N_g>$ - $N_h$, and $<N_b>$ - $N_h$. The lines in the figure are fitted results and are given by $<N_s> = 0.56N_g ± 0.57$, $<N_s> = 0.78N_b ± 0.42$, $<N_s> = 0.79N_h ± 0.28$, $<N_g> = 0.74N_b ± 0.45$, $<N_g> = 0.98N_h ± 0.17$ and $<N_b> = 0.95N_h ± 0.19$. Thus, we can see that $<N_s>$ increases with the increases of $N_g$, $N_b$ and $N_h$ and then drops at high multiplicity. The behavior of $<N_s>$ at high multiplicity reflects a saturation phenomenon accompanying light projectile nucleus due to the stopping power of target nucleus. The saturation value reflects the maximum $N_s$ value at ~1 A GeV [14, 15]. The figure 2(e) and 2(f) shows that $<N_g>$ and $<N_b>$ increase with the increase of $<N_h>$. We have not observed the saturation phenomenon for $<N_b>$ because the target spectator has enough size to evaporate black fragments in collisions induced by light projectile.

## 5. Conclusions

In the present article, we studied the characteristics of the multiplicity distribution of shower particles and target fragments by using multi-component Erlang distribution of multi-source thermal model. The multiplicity distributions of shower particles and grey fragments are in a narrow range and appear to be relatively smooth curves. The multiplicity distributions of black fragments and heavily ionization fragments are in a wide range and appear with relative large fluctuations due to low statistics. The distributions of four kinds of multiplicities are well described by two-component Erlang distribution having approximately equal weight for each component.

In the multiplicity correlations, the behavior of shower particle multiplicity reflects a saturation phenomenon with light projectile nucleus due to the stopping power of the target nucleus. In case of grey and black fragments there is no saturation phenomenon observed because the target has enough size to support cascade collisions and evaporation processes.

**Acknowledgement:** Authors are grateful to the all technical staff of GSI, Germany for exposing nuclear emulsion detector with $^{84}Kr_{36}$ beam.

**Table 1:** Parameter values and the corresponding $\chi^2$/dof for the curves in Figure 1.

| Figure | $<\alpha_{i1}>$ | $\beta_1$ | $k_1$ | $<\alpha_{i2}>$ | $\beta_2$ | $\chi^2$/dof |
|---|---|---|---|---|---|---|
| 1(a) | 0.8 | 3 | 0.50 | 1.2 | 3 | 0.55 |
| 1(b) | 1.2 | 1 | 0.47 | 1.8 | 2 | 0.57 |
| 1(c) | 0.8 | 2 | 0.52 | 1.7 | 3 | 0.83 |
| 1(d) | 1.3 | 3 | 0.57 | 1.9 | 4 | 0.89 |

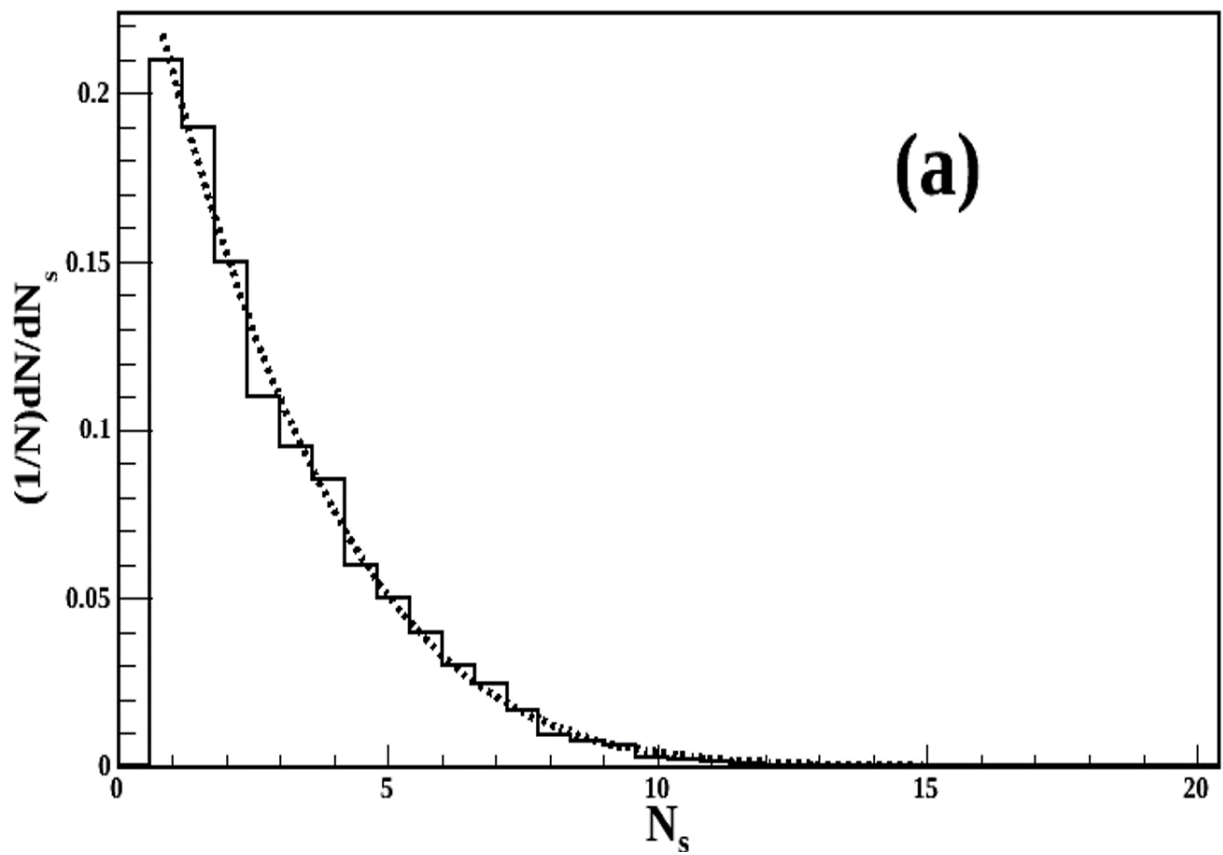

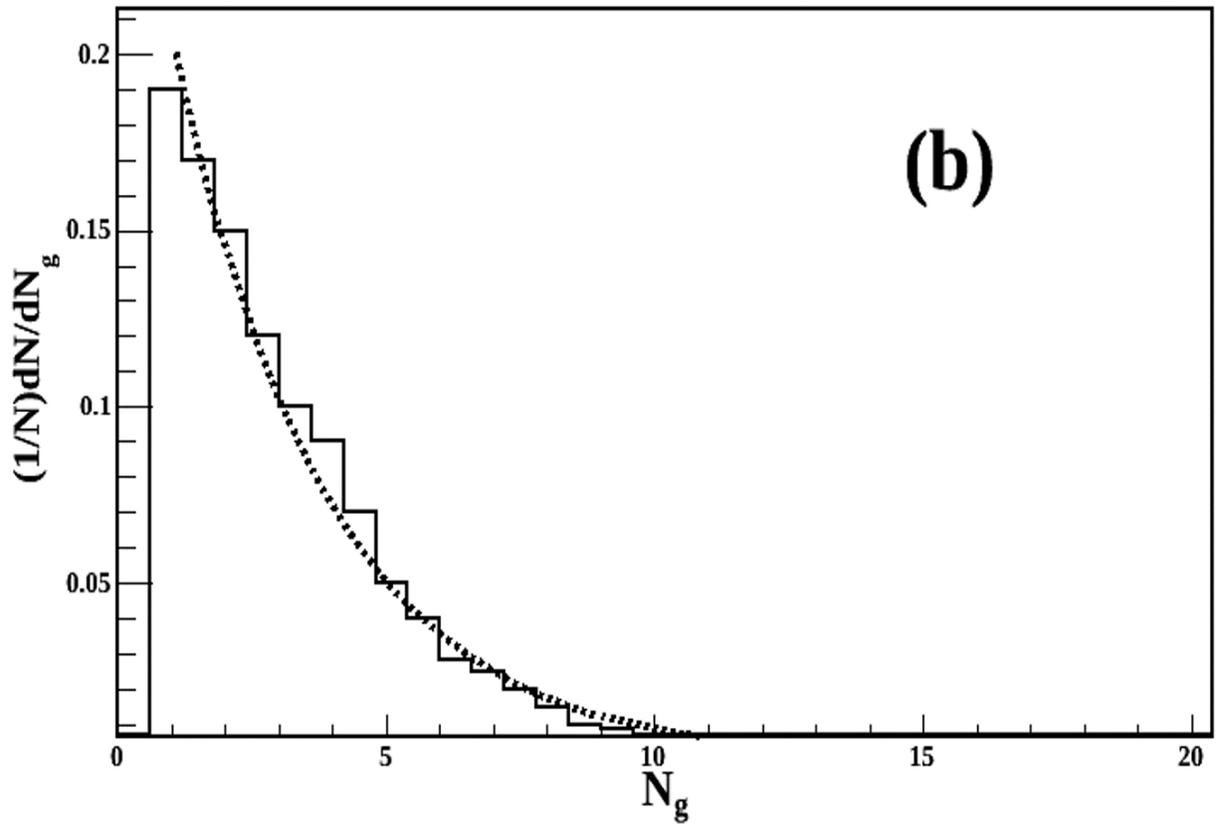

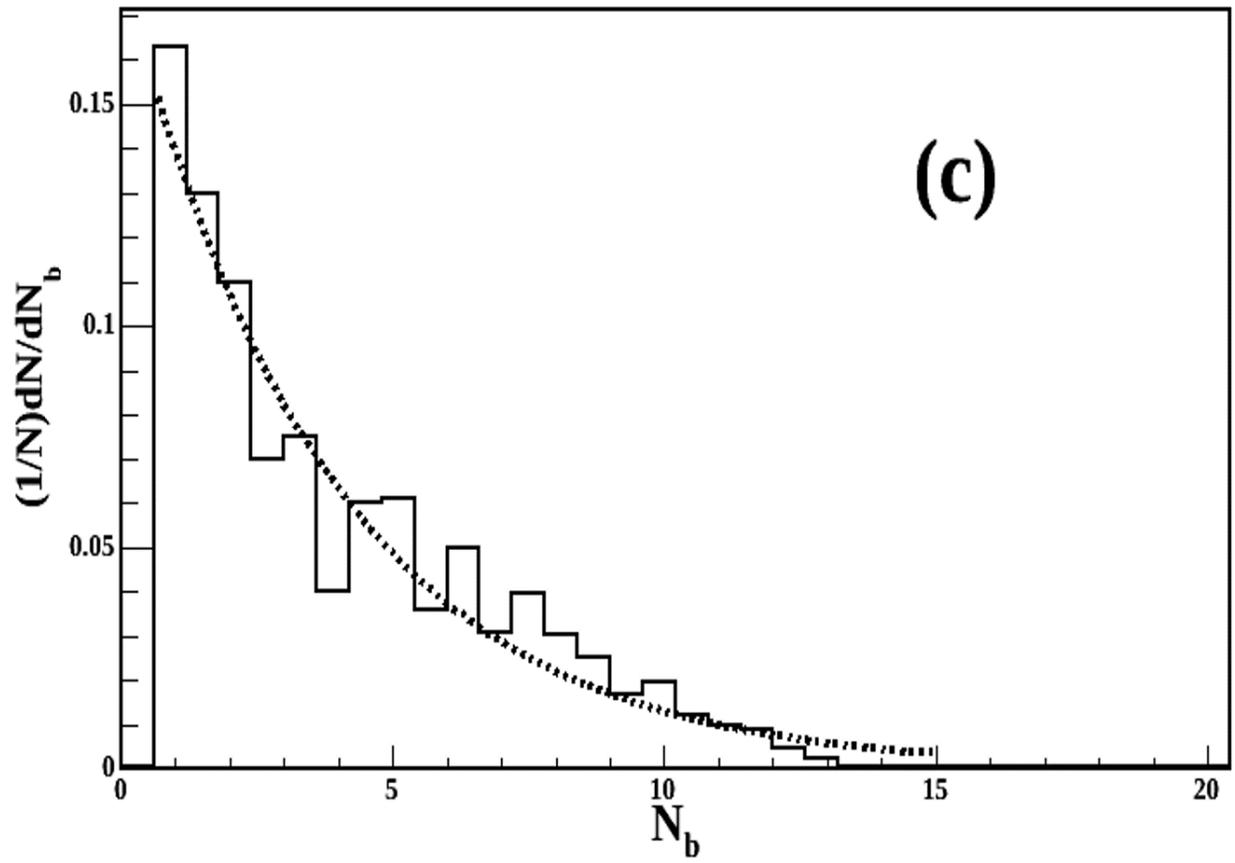

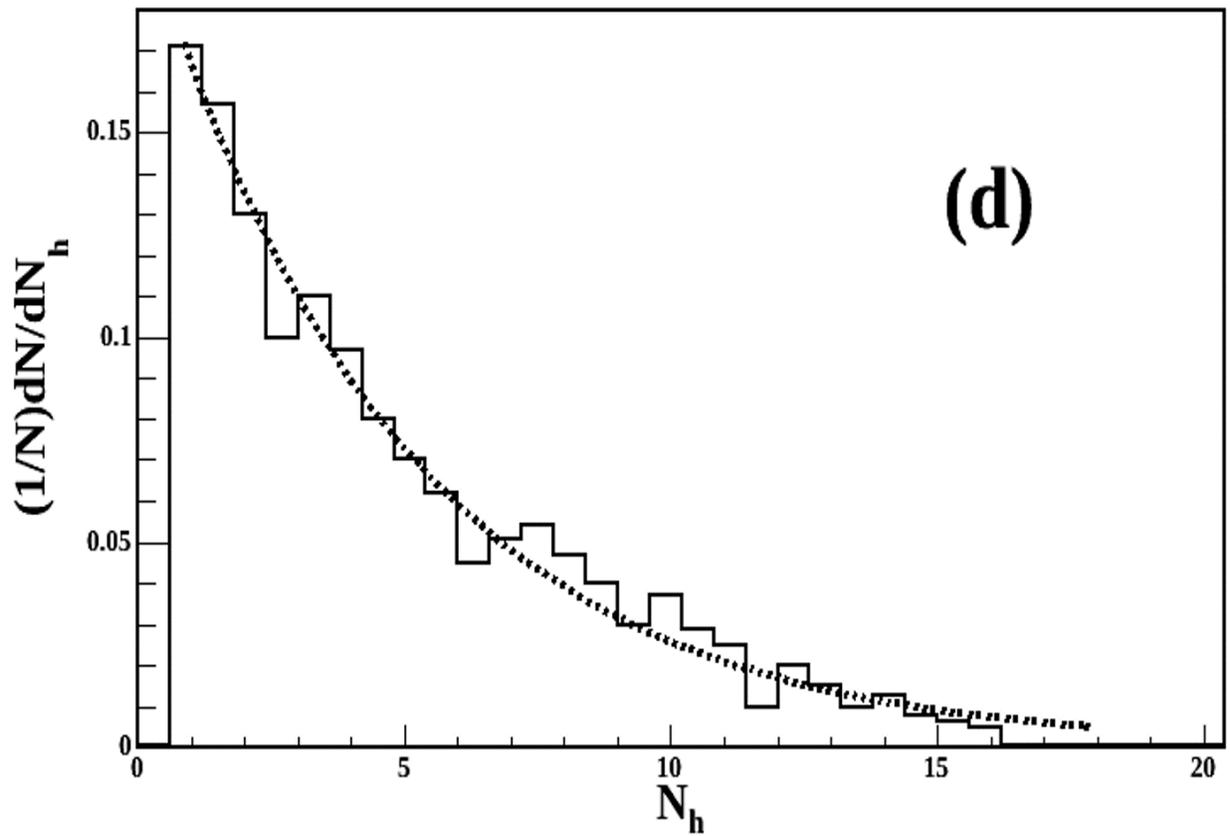

**Fig. 1:** Multiplicity distributions of (a) shower particles, (b) grey fragments, (c) black fragments, and (d) heavily ionization fragments produced in $^{84}Kr_{36}$ – Emulsion Interactions at ~1 GeV per nucleon. The histograms are experimental data and dotted curves are results from the model.

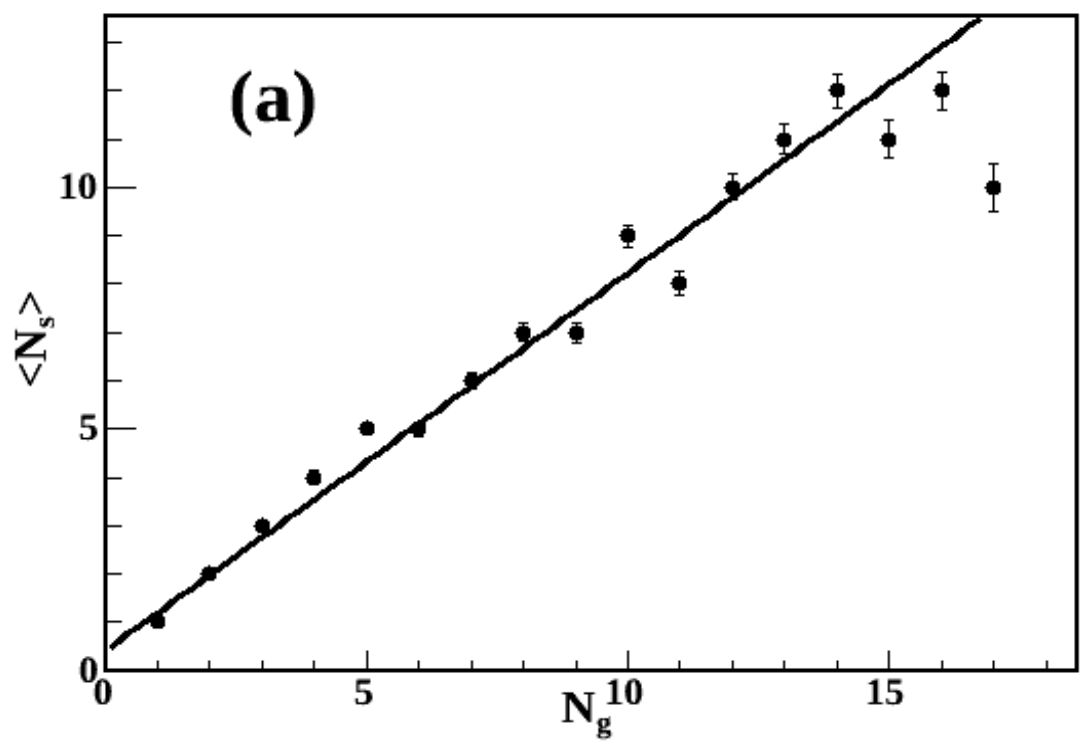

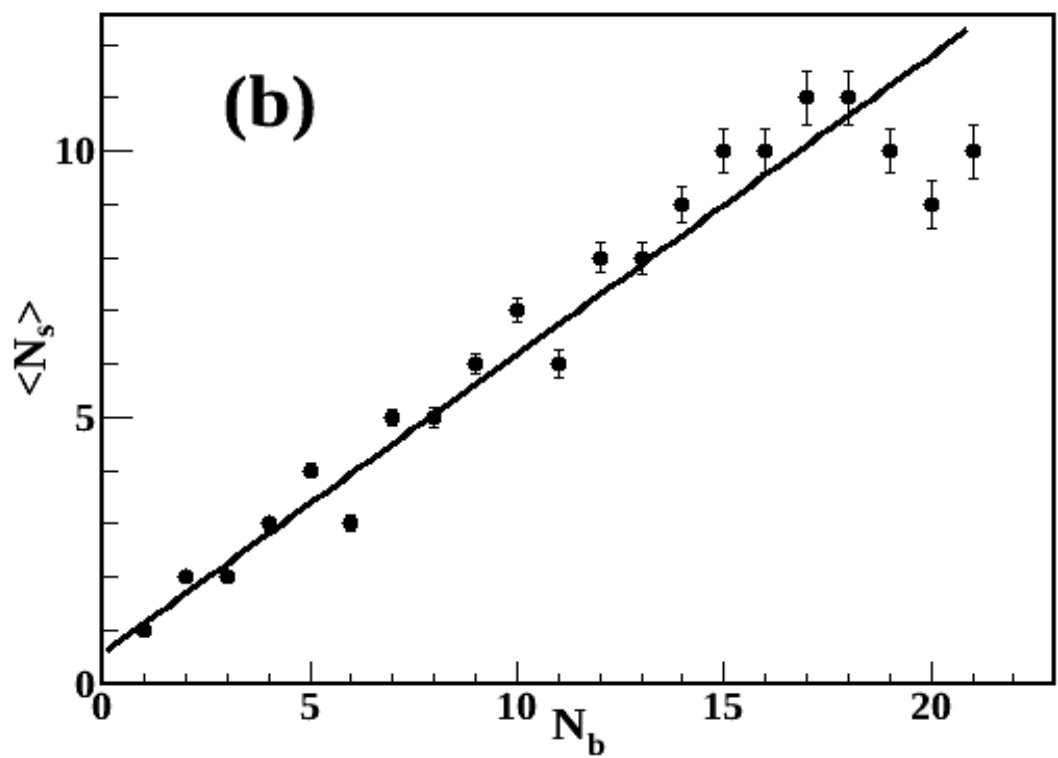

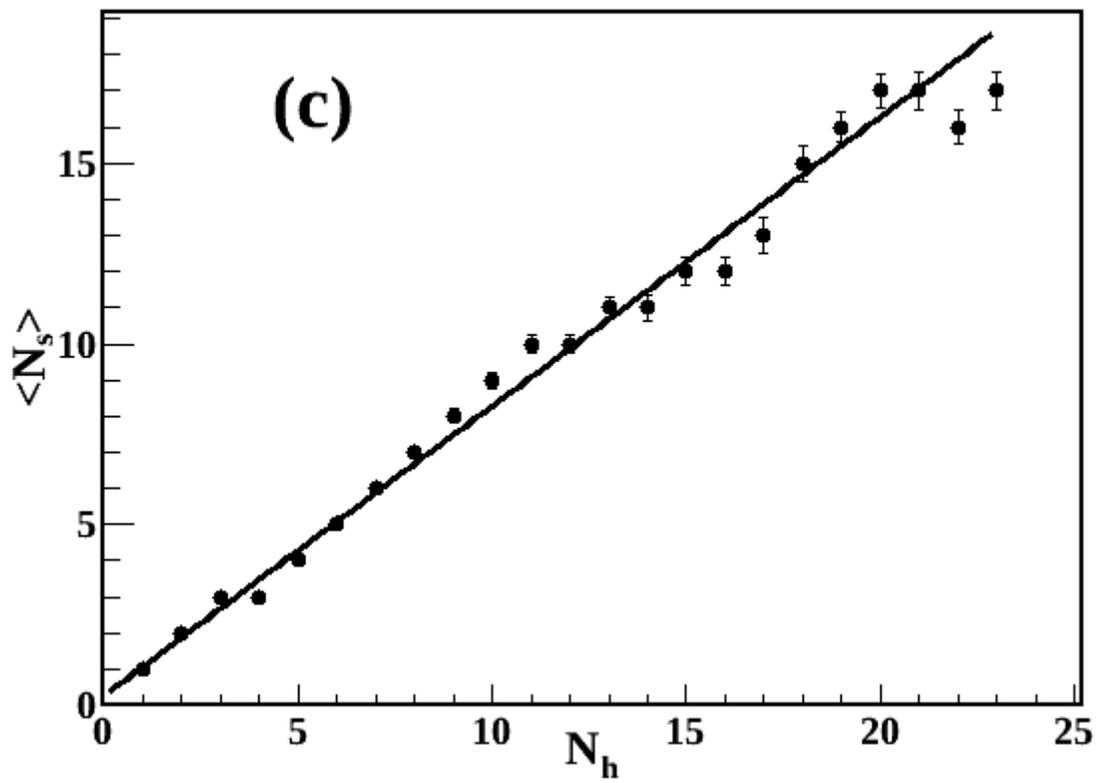

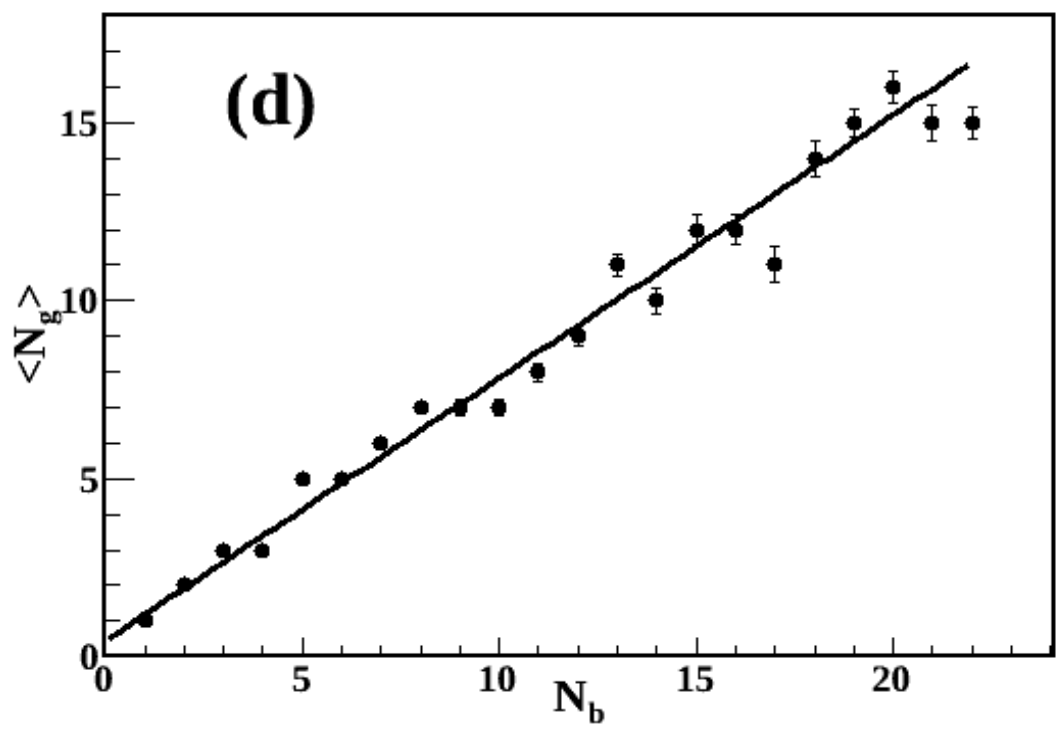

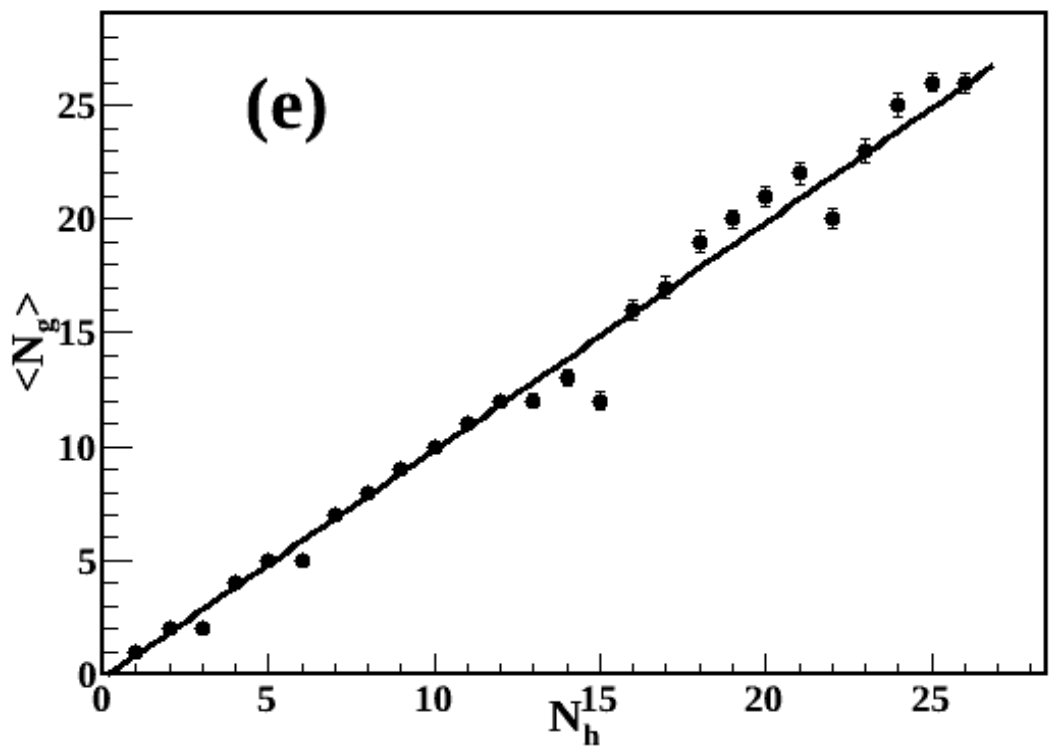

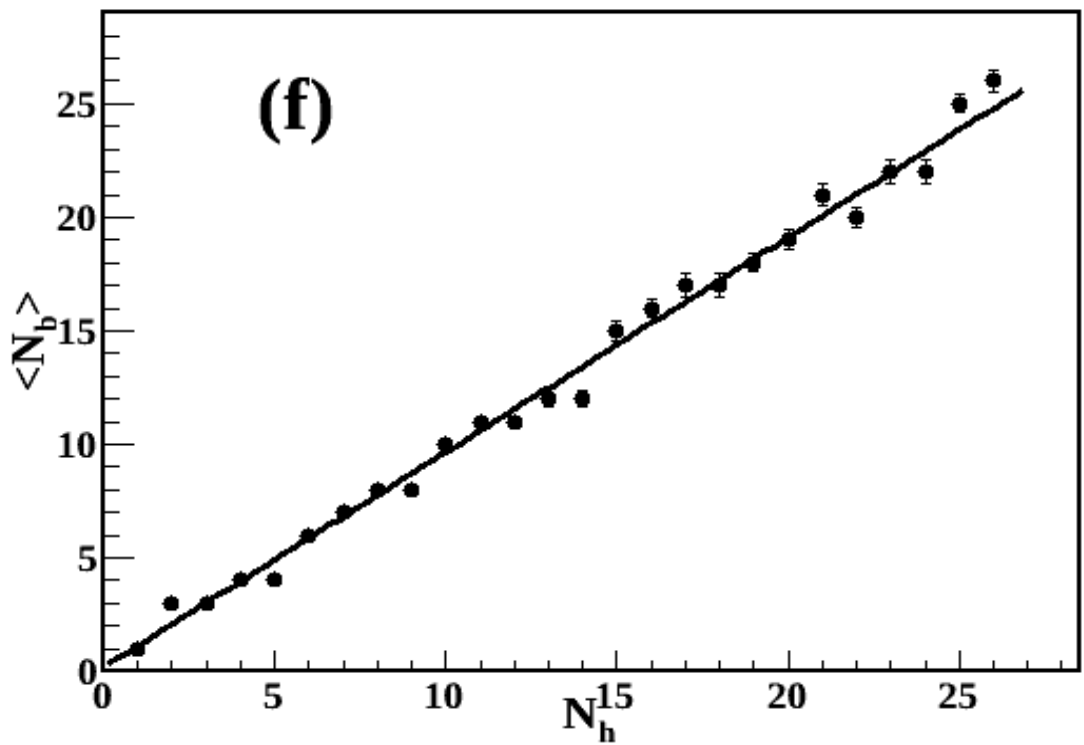

**Fig. 2:** Correlations between (a) $<N_s>$ - $N_g$, (b) $<N_s>$ - $N_b$, (c) $<N_s>$ - $N_h$, (d) $<N_g>$ - $N_b$, (e) $<N_g>$ - $N_h$, and (f) $<N_b>$ - $N_h$ in $^{84}Kr_{36}$ - Emulsion Interactions at ~1 A GeV per nucleon. The points are our experimental data and the lines are the fitted results.